\documentclass{elsart}

\usepackage{amsfonts}

\begin{document}
\begin{frontmatter}
\title{Extending $SU(2)$ to $SU(N)$ QCD.}
\author{M.L. Walker\thanksref{JSPS}\thanksref{Nihon}}
\address{College of Arts and Sciences, Chiba University, Inage-ku, Chiba, Japan 263-8522}
\thanks[JSPS]{Supported by a fellowship from the Japan Society for the Promotion 
of Science (P05717)}
\thanks[Nihon]{now at Institute of Quantum Science, College of Science and Technology, 
Nihon University, Chiyoda-ku, Tokyo, Japan 101-8308}
\date{}
\begin{keyword}
QCD, Abelian dominance
\end{keyword}

\begin{abstract}
Abelian dominance is used to reformulate the QCD Lagrangian
as a sum over the roots of Lie group representation theory. 
This greatly facilitates extending the $SU(2)$ magnetic ground state energy spectrum, 
several arguments for the stability of the magnetic ground state,
and the Faddeev-Skyrme model to arbitrary $SU(N)$ QCD. 
\end{abstract}

\end{frontmatter}

\section{Introduction}
By far the majority of significant analyses of QCD are first performed in two-colour QCD 
for the sake of mathematical simplicity. However such a calculation can only be regarded
as a toy until the equivalent calculation is performed for three or more colours.
A principle or technique that allowed for straightforward extension of two-colour results
to QCD with three or more colours would constitute a powerful time-saver.

This paper uses Lie algebra representation theory
to express $N$-colour QCD as a sum over copies
of two-colour QCD. The resulting expression neglects the interactions between
off-diagonal gluons of different root vectors. it can be argued however that this truncation
follows from Abelian dominance, for which there has long been both analytic and
numerical evidence \cite{diG98,P97,BOT97,SNW94,C00,K98a}. This 
makes it easy to extend calculations for two-colour QCD to QCD with arbitrarily many colours. 

Abelian dominance states that the dynamics of QCD, especially in the low-energy limit,
are dominated by the Abelian and monopole components of the gluon field. Assuming Abelian 
dominance, quark confinement by electric vortices can be proven \cite{EI82}, but this paper
is more concerned with the dynamics of the off-diagonal gluons which lie at the centre of 
the monopole condensate stability issue. One expects from Abelian dominance that the 
dominant interactions are with the Abelian and monopole gluon components, so those interactions 
are retained along with self-interaction terms that lie in the Abelian direction since 
they couple to the Abelian component of the field strength. However other
self-interactions, \textit{ie.} interactions between off-diagonal
gluons corresponding to different root vectors, are neglected.

The Abelian component is specified by the Cho-Faddeev-Niemi (CFN) decomposition 
\cite{Cho80b,C80,FN99b}, a gauge invariant way of specifying the Abelian dynamics 
and topological component of QCD. Employing the root vector notation of $SU(N)$ representation theory makes it possible to express the $SU(N)$ QCD Lagrangian as a sum over copies of the
two-colour theory. Both of these are explained in section \ref{sec:CFN}. 
Section \ref{sec:apps} applies this formalism to the
ground-state energy spectrum and analysis of the alleged unstable mode.
The final topic for consideration is the Faddeev-Skyrme model which is
found to extend to higher colours very easily. Section \ref{sec:discuss} is a summary.

\section{Representing the Gluon Field} \label{sec:CFN}

The CFN decomposition \cite{Cho80a,FN99c} is used to specify the
Abelian components of the background field in a gauge invariant manner. This is an
improvement over the maximal Abelian gauge \cite{tH81}, in which many of the
results discussed in this paper were originally derived. 
However repeating them in the CFN decomposition,
or alternately reducing the CFN decomposition to the maximal Abelian gauge by 
gauging all the Abelian directions to be constant in spacetime, is trivial.

The Lie group $SU(N)$ has $N^2-1$ generators $\lambda^{(i)}$, of which $N-1$
are Abelian generators $\Lambda^{(i)}$. 
For simplicity, the gauge transformed Abelian directions (Cartan generators)
are denoted
\begin{equation}
\hat{n}_i = U^\dagger \Lambda^{(i)} U. 
\end{equation}
In the same way, the standard
raising and lowering operators for the weights $E_{\pm\alpha}$ with the gauge 
are replaced by the transformed 
\begin{equation}
E_{\pm \alpha} \rightarrow U^\dagger E_{\pm \alpha} U,
\end{equation}
where $E_{\pm \alpha}$ refers to the gauge transformed operator 
throughout the rest of this article.

Gluon fluctuations in the $\hat{n}_i$ directions are described by $c^{(i)}_\mu$. 
There is a covariant derivative which leaves
the $\hat{n}_i$ invariant, 
\begin{eqnarray}
\hat{D}_\mu \hat{n}_i \equiv (\partial_\mu + g\vec{V}_\mu \times) \hat{n}_i = 0,
\end{eqnarray}
where $\vec{V}_\mu$ is of the form
\begin{eqnarray}
\vec{V}_\mu &=& c^{(i)}_\mu \hat{n}_i + \vec{B}_\mu , \nonumber \\
\vec{B}_\mu &=& g^{-1} \partial_\mu \hat{n}_i \times \hat{n}_i,
\end{eqnarray}
and summation is implied over $i$. $\vec{X}_\mu$ denotes the dynamical 
degrees of freedom (DOF) perpendicular to $\hat{n}_i$, 
so if $\vec{A}_\mu$ is the gluon field then
\begin{eqnarray}
\vec{A}_\mu &=& \vec{V}_\mu + \vec{X}_\mu \nonumber \\
&=& c^{(i)}_\mu \hat{n}_i + \vec{B}_\mu + \vec{X}_\mu,
\end{eqnarray}
where
\begin{eqnarray}
\vec{X}_\mu &\bot& \hat{n}_i \nonumber \\
\vec{X}_\mu &=& g^{-1}\hat{n}_i \times \vec{D}_\mu \hat{n}_i, \nonumber \\
\vec{D}_\mu &=& \partial_\mu + g\vec{A}_\mu \times. \nonumber \\
\end{eqnarray}

The field strength tensor of QCD expressed in terms of the CFN decomposition is
\begin{eqnarray}
\vec{F}_{\mu\nu} &=& (\partial_\mu c^{(i)}_\nu - \partial_\nu c^{(i)}_\mu)\hat{n}_i 
+(\partial_\mu \vec{B}_\nu - \partial_\nu \vec{B}_\mu + g \vec{B}_\mu \times \vec{B}_\nu) \nonumber \\
&& + (\mathbf{\hat{D}}_\mu \vec{X}_\nu - \mathbf{\hat{D}}_\nu \vec{X}_\mu)
+g \vec{X}_\mu \times \vec{X}_\nu .
\end{eqnarray}

Because $\vec{X}_\mu$ is orthogonal to all Abelian directions it can be expressed as 
a linear combination of the raising and lowering operators $E_{\pm\alpha}$, which
leads to the definition
\begin{eqnarray}
X_\mu^{(\pm \alpha)} \equiv E_{\pm\alpha} \mbox{Tr}[\vec{X}_\mu E_{\pm\alpha}],
\end{eqnarray}
so
\begin{equation}
X_\mu^{(-\alpha)} = {X_\mu^{(+ \alpha)}}^\dagger .
\end{equation}

Restricting the interaction terms to those that couple to Abelian fields, the
field strength tensor becomes
\begin{eqnarray} \label{eq:fieldstrength}
\vec{F}_{\mu\nu} &=& \sum_{\alpha>0} \Big[
\alpha^{(i)} \sqrt{\frac{2}{N}}(\partial_\mu c^{(i)}_\nu - \partial_\nu c^{(i)}_\mu) \hat{n}_i \nonumber \\
&& +\sqrt{\frac{2}{N}}(\partial_\mu \vec{B}^{(\alpha)}_\nu - \partial_\nu \vec{B}^{(\alpha)}_\mu 
+ g \vec{B}^{(\alpha)}_\mu \times \vec{B}^{(\alpha)}_\nu) \nonumber \\
&& + (\mathbf{\hat{D}}^{(\alpha)}_\mu \vec{X}^{(\alpha)}_\nu 
- \mathbf{\hat{D}}^{(\alpha)}_\nu \vec{X}^{(\alpha)}_\mu)
+g \vec{X}^{(+\alpha)}_\mu \times \vec{X}_\nu^{(-\alpha)} \Big],
\end{eqnarray}
where $\vec{B}^{(\alpha)}_\mu$ represents the monopole
fields felt by the valence gluon $\vec{X}_\nu^{(\alpha)}$.

Cross terms between $\vec{X}^{(\alpha)}_\mu$ of different root vectors 
$\vec{\alpha}$ have clearly been neglected. These do not lie in the 
Abelian direction and do not
couple to the Abelian field and are therefore expected to be of minimal importance
according to Abelian dominance. The self interaction
$\vec{X}^{(+\alpha)}_\mu \times \vec{X}_\nu^{(-\alpha)}$ by contrast, does lie in the
Abelian direction and is therefore expected to contribute significantly to
the $\vec{X}_\mu$ dynamics at low energies. Even the four-point self-interaction
\begin{displaymath}
\left(\vec{X}^{(+\alpha)}_\mu \times \vec{X}_\nu^{(-\alpha)}\right)^2,
\end{displaymath}
can receive corrections from the Abelian dynamics. Indeed, it has already been
argued~\cite{KKP05,F83} that this four-point term plays an essential role in stabilising
the monopole condensate. This is discussed in greater detail in subsection \ref{subsec:quartic}.

The first term on the second last line of (\ref{eq:fieldstrength}) 
contains the $\vec{X}_\mu$ derivatives which do not
constitute Abelian dynamics. They have been retained to give the $\vec{X}_\mu$ a propagator,
since the off-diagonal dynamics are of interest. It was also possible to use
an auxiliary field similar to \cite{K98a}, but this formalism maintains an advantageous 
resemblance to two-colour QCD.

The corresponding Lagrangian is 
\begin{eqnarray} \label{eq:oneloop}  
\mathcal{L} &=& -\frac{1}{4}\sum_{\alpha>0} \Big[
\frac{2}{N}(\partial_\mu c^{(\alpha)}_\nu - \partial_\nu c^{(\alpha)}_\mu)^2 \nonumber \\
&&+\frac{2}{N}(\partial_\mu \vec{B}^{(\alpha)}_\nu - \partial_\nu \vec{B}^{(\alpha)}_\mu 
+ g \vec{B}^{(\alpha)}_\mu \times \vec{B}^{(\alpha)}_\nu)^2 \nonumber \\
&&+ \frac{4}{N}(\partial_\mu c^{(\alpha)}_\nu - \partial_\nu c^{(\alpha)}_\mu)
\hat{n}_\alpha \cdot (\partial_\mu \vec{B}^{(\alpha)}_\nu - \partial_\nu \vec{B}^{(\alpha)}_\mu 
+ g \vec{B}^{(\alpha)}_\mu \times \vec{B}^{(\alpha)}_\nu) \nonumber \\
&&+2g(\partial_\mu c^{(\alpha)}_\nu - \partial_\nu c^{(\alpha)}_\mu)
\hat{n}_\alpha \cdot (\vec{X}^{(+\alpha)}_\mu \times \vec{X}^{(-\alpha)}_\nu) \nonumber \\
&&+ 2g(\partial_\mu \vec{B}^{(\alpha)}_\nu - \partial_\nu \vec{B}^{(\alpha)}_\mu 
+ g \vec{B}_\mu^{(\alpha)} \times \vec{B}_\nu^{(\alpha)}) \cdot
(\vec{X}^{(+\alpha)}_\mu \times \vec{X}^{(-\alpha)}_\nu) \nonumber \\
&&+ (\mathbf{\hat{D}}_\mu^{(\alpha)} \vec{X}^{(\alpha)}_\nu 
- \mathbf{\hat{D}}^{(\alpha)}_\nu \vec{X}^{(\alpha)}_\mu)^2 
+ g^2 (\vec{X}_\mu^{(+\alpha)} \times \vec{X}_\nu^{(-\alpha)})^2 \Big] \nonumber \\
&& -\frac{1}{4}\sum_{\stackrel{\alpha>\beta>0}{\alpha\cdot \beta = \half}}
g^2 (\vec{X}_\mu^{(+\alpha)} \times \vec{X}_\nu^{(-\alpha)})
\cdot (\vec{X}_\mu^{(+\beta)} \times \vec{X}_\nu^{(-\beta)})
\end{eqnarray}
With regard to gauge-fixing, the CFN decomposition leaves QCD with an extended gauge
symmetry $SU(N) \otimes (SU(N)/U(1)^{N-1})$, where the additional freedom comes
from the ability to rotate the $\hat{n}_i$. To avoid distraction from the main point of
this article I simply state here that this and related issues regarding the
interpretation of the CFN decomposition have been thoroughly discussed and 
resolved \cite{S99,CP02,BCK02,KMS05,K06}.
With the essential dynamics expressed as a sum over $SU(2)$ dynamics, it is now
convenient to add gauge-fixing and ghost terms which can be copied directly from
two-colour QCD. They are
\begin{eqnarray} 
\mathcal{L}_{ghost}^{one-loop} &=& \sum_{\alpha>0} \Big[
- \frac{1}{\xi_1}|\hat{D}^{(\alpha)}_\mu \vec{X}_\mu^{(\alpha)}|^2
- i\bar{\mathbb{C}}^{(\alpha)}_\theta \cdot 
\hat{D}^{(\alpha)}_\mu \hat{D}^{(\alpha)}_\mu \mathbb{C}_\theta^{(\alpha)} \nonumber \\
&&-\frac{1}{\xi_2} (\partial_\mu \vec{A}^{(\alpha)}_\mu)^2
+ i\bar{\mathbb{C}}^{(\alpha)}_\omega 
\cdot \partial_\mu \vec{D}^{(\alpha)}_\mu \mathbb{C}^{(\alpha)}_\omega \Big],
\end{eqnarray}
to first loop order. Conventional gauge fixing is given by the second line in this equation.
The first line restricts the additional gauge symmetry
introduced by using the CFN decomposition. It is greatly simplified at the one-loop 
approximation. The full gauge-fixing/ghost Lagrangian is
\begin{eqnarray} \label{eq:gaugefix}
\mathcal{L}_{ghost} = \sum_{\alpha>0} \Big[
- \frac{1}{\xi_1}|\hat{D}^{(\alpha)}_\mu \vec{X}_\mu^{(\alpha)}|^2 
-\frac{1}{\xi_2} (\partial_\mu \vec{A}_\mu^{(\alpha)})^2
+ i\bar{\mathbb{C}}^{(\alpha)}_\omega 
\cdot \partial_\mu \vec{D}^{(\alpha)}_\mu \mathbb{C}^{(\alpha)}_\omega 
\nonumber \\
- i\bar{\mathbb{C}}^{(\alpha)}_\theta 
\cdot (\partial_\mu + g(\vec{V}_\mu - \vec{X}_\mu)^{(\alpha)}\times)
(\partial_\mu + g(\vec{V}_\mu + \vec{X}_\mu)^{(\alpha)}\times)
\mathbb{C}_\theta^{(\alpha)}\Big], 
\end{eqnarray}
A full derivation explains
the interpretation of the CFN decomposition which is well-beyond the scope of
this article. The conventions in this paper are based on \cite{KMS05}.

The sum over positive roots is more than a convenient shorthand. It 
indicates that assuming Abelian dominance almost reduces
the dynamics of $N$-colour QCD to multiple copies of $SU(2)$. The discrepency is
the last line of (\ref{eq:oneloop}) which contains all the cross terms between different
root vectors.  This was
the one-loop finding of Cho \textit{et.~al.} \cite{CKP05} in their analysis of the monopole
condensate's stability in three-colour QCD. Indeed, their expression of the 
one-loop
Lagrangian is a special case of (\ref{eq:oneloop}), because the quartic interactions 
do not contribute to their calculation at one-loop. This makes it easy to find
any low-energy result in $N$-colour QCD if the corresponding result is known
for the two-colour theory and the mixed quartic terms can be neglected, as is 
generally the case at low loop order.

\section{An Extension of Two-Colour Results to Higher Colours} \label{sec:apps}

\subsection{QCD Magnetic Vacuum}
There is an additional factor of $\frac{2}{N}$ in front of the Abelian
field strength but not in front of the other terms in (\ref{eq:oneloop}).
It is not difficult to see that substituting this into the derivation of the
one-loop order calculation of the ground state energy of the magnetic background 
will yield the long known \cite{S77,F80} one-loop expression for the 
energy of the background magnetic field strength $H$ in
$N$-colour QCD, namely
\begin{eqnarray} \label{eq:NSavvidy}
\mathcal{H}_{SU(N)} &=& \sum_{(\alpha > 0)} \left(H^{(\alpha)^2} 
\left(\frac{2}{N} \frac{1}{2g^2} 
+ \frac{11}{48\pi^2} \ln \frac{H^{(\alpha)}}{\mu^2} \right) \right) \nonumber \\
&=& \sum_{(\alpha > 0)} \left(H^{(\alpha)^2} \left(\frac{1}{Ng^2} 
+ \frac{11}{48\pi^2} \ln \frac{H^{(\alpha)}}{\mu^2} \right) \right).
\end{eqnarray}
The imaginary part has been neglected due to the long-running
controversy regarding its value and physical interpretation. It is
discussed in the following subsections.

\subsection{Zero-point Eigenvalue Spectrum} \label{subsec:eigenvalue}
Assuming a covariant constant
background and keeping only quadratic terms it is straightforward to find
that the energy eigenvalue spectrum of each $\vec{X}^{(\alpha)}_\mu$
by any approach used to find that of two-colour QCD. If $H^{(\alpha)},
E^{(\alpha)}$ are the magnetic and electric backgrounds respectively felt
by $\vec{X}^{(\alpha)}_\mu$, then its energy eigenvalues are
\begin{eqnarray}
 \lambda &=& 2gH^{(\alpha)}((n+1)\pm 1/2)+2gE^{(\alpha)}((m+1)\mp 1/2) ,
\nonumber \\
 \lambda &=& 2gH^{(\alpha)}(n\pm 1/2)+2gE^{(\alpha)}(m\mp 1/2) ,
\end{eqnarray}
where $n,m=0,1,2,\cdots$. 

In a pure magnetic background each $\vec{X}_\mu^{(\alpha)}$ has
an $n=0$ mode that contributes a destabilising 
imaginary part. This 
has lead many to believe that the Savvidy background is unstable. From 
Chang and Weiss' \cite{CW79} analysis of the unstable modes in $SU(2)$ QCD 
it follows that the total density of unstable modes is 
\begin{equation}
\pi^{-2}\sum_{\alpha >0} \left(gH^{(\alpha)}\right)^\frac{3}{2},
\end{equation}
where there are $\frac{N^2-N}{2}$ $\vec{X}^{(\alpha)}$ contributing.

There have however been several claims that this imaginary component is
an artifact of the quadratic approximation and/or renormalisation scheme.
These are discussed subsequently.

The eigenvalues are all non-negative only in the
self-dual case $H^{(j)}=E^{(j)}$. Then the eigenvalues read
\begin{eqnarray}
 \lambda  &=& 2gH^{(j)}(n+m+2) > 0 , \nonumber \\
 \lambda  &=& 2gH^{(j)}(n+m)  \ge 0 \; (n=m=0: \textit{zero mode}),
\end{eqnarray}

The eigenvalues for the (off-diagonal) ghosts, are
\begin{eqnarray}
\lambda_{n,m}' = 2gH^{(j)}(n+1/2)+2gE^{(j)}(m+1/2) > 0 ,
\end{eqnarray}
which is always positive. 
Note that the Gribov problem does not manifest at one-loop. 

\subsection{Renormalisation by Causality}
Cho and Pak \cite{CP02} argued that the imaginary part was an artifact of 
zeta-function renormalisation being too naive.
Their 'renormalisation by causality' at one-loop found no imaginary part 
for the magnetic background but it did find one
for the electric background, as confirmed by
a perturbative calculation by the current author \cite{CmeP04,Cme04}.
Together with Kim \cite{CKP05} they extended this result to three-colour QCD by 
also expressing the Lagrangian as a sum of two-colour theories, although
root vectors were not used explicitly, and derived 
the one-loop effective theory for three-colour QCD. Their results, as they note,
extend easily to $N$-colour QCD. The imaginary part of its action is a 
sum over the $\frac{N^2-N}{2}$ copies of two-colour QCD,
\begin{eqnarray} \label{eq:imag}
&0& \hspace{1cm} \mbox{pure magnetic background} \nonumber \\
&-\frac{11g^2}{96}\sum_{\alpha >0} {E^{(\alpha)}}^2 &\hspace{1cm} \mbox{pure electric background}
\end{eqnarray}
where $E^{(\alpha)}$ is the strength of the
electric background felt by $\vec{X}_\mu^{(\alpha)}$. This is physically 
interpreted as the magnetic background being stable but the electric 
background decaying by the annihilation of gluons \cite{CP02}.

\subsection{Stability Through Effective Gluon Mass?} \label{subsec:gluonmass}
Kondo \cite{K04} argued that the imaginary contribution to the energy eigenvalues could
be removed by an effective gluon mass, so the eigenvalue equation looks like
\begin{equation}\label{eq:massspectrum}
\lambda = \sqrt{k^2 + M^2 + gH\left(n-\half \right)},\; n=0,1,2,\ldots
\end{equation}
His effective mass term came from the quartic term
\begin{equation}
g^2 (\vec{B}_\mu \times \vec{B}_\nu) \cdot (\vec{X}_\mu \times \vec{X}_\nu).
\end{equation}
(He has also constructed an analogous argument based on the condensate
$\langle -\vec{X}_\mu \vec{X}_\mu \rangle$ which is discussed in subsection
\ref{subsec:quartic}.) He diagonalised the mass term 
\begin{equation}
(M_X^2)^{ad} = g^2\epsilon^{abc} \epsilon^{dec} B_\rho^b B_\rho^e,
\end{equation}
to find the non-zero mass eigenvalues 
\begin{equation}\label{eq:gluonmass}
M_X^2 = g^2\vec{B}_\rho \cdot \vec{B}_\rho,
\end{equation}
and derived the result
\begin{equation}
M_X^2 \ge \sqrt{2}|gH|,
\end{equation}
which is sufficient to provide the stability in $SU(2)$ QCD. In $SU(N)$ QCD
the corresponding quartic terms are
\begin{equation}
g^2 (\vec{B}_\mu^{(\alpha)} \times \vec{B}_\nu^{(\alpha)}) \cdot
(\vec{X}^{(+\alpha)}_\mu \times \vec{X}^{(-\alpha)}_\nu).
\end{equation}
By the above reasoning the off-diagonal gluon $\vec{X}^{(\alpha)}$ gains 
an effective mass squared of 
\begin{equation}
{M_X^{(\alpha)}}^2 = g^2\vec{B}^{(\alpha)}_\rho \cdot \vec{B}^{(\alpha)}_\rho
\ge \sqrt{2}|gH^{(\alpha)}|.
\end{equation}
An alternative adaptation of this argument to $N=3,4$ QCD is presented
in \cite{me07}.

Such arguments can be seen as
requiring an imposed mechanism, although it is my view that they illustrate that the 
condensate indicated by the effective energy stabilises itself. The point
could still be made that mass cannot be calculated
quantum mechanically from first principles. There would be issues with gauge invariance
if it could. However there is a very simple argument based on fundamental principles why
an object of zero or very small mass should acquire dynamical mass when it is confined.
When an object is confined its de Broglie wavelength is automatically bound to be less than
the confinement length. This puts a lower limit on the non-zero energy spectrum. 
Such gaps in the energy spectrum are exactly 
what studies of mass generation consider to be the indicator of mass. This hand-waving
connection between confinement and mass generation is consistent with numerical
deconfinement studies which find that the critical temperatures for
deconfinement and chiral symmetry breaking are well-defined and identical 
for low mass quarks and quarks, but that
the transition becomes a cross-over for heavy bare quarks 
(see \cite{HF04,KLP01,YS82,PW84} and references therein).

\subsection{Quartic Terms of Unstable Modes} \label{subsec:quartic}
An argument for condensate stability was made by Flory \cite{F83b}, and again later
by Kay, Kumar and Parthasarathy 
\cite{KKP05}, who demonstrated that including the quartic terms 
\begin{eqnarray} \label{eq:quartic}
-\frac{1}{4}g^2\sum_{\alpha>0} 
(\vec{X}^{(+\alpha)}_\mu \times \vec{X}^{(-\alpha)}_\nu)^2 ,
\end{eqnarray}
related to the 
unstable modes removed the imaginary part from the effective action
and confirmed the real part of the effective action of two colour QCD.
It is straightforward to adapt the calculations in \cite{KKP05,F83} and find
that the imaginary part is removed by these quartic terms, 
and the real part of the effective action is confirmed.

As those authors noted, the original calculation
neglects the quartic terms mixing the unstable modes with the stable ones. For $N>2$ 
the quartic terms mixing $\vec{X}^{(\alpha)}_\mu$ 
of different root vectors, the final line in eq.~(\ref{eq:oneloop}) are also neglected.
It seems unlikely however, that they would undo the stabilising effect of (\ref{eq:quartic})
since they have the same sign.
Other mixed quartic terms are excluded automatically by Abelian dominance. 
Cubic terms don't contribute. \cite{KKP05,F83}

The quartic term featured in (\ref{eq:quartic}) was the crux of
another argument in two-colour QCD
by Kondo \cite{K06} for off-diagonal mass generation. He 
demonstrated that the condensate $\langle -\vec{X}_\rho \vec{X}_\rho \rangle \ne 0$, which
when substituted into the Lagrangian yields a gluon mass term similar to
that in subsection \ref{subsec:gluonmass}. This argument also 
follows for arbitrarily many colours where it is augmented by the quadratic cross-terms.

\subsection{Faddeev-Skyrme Model} \label{sec:skyrme}
An off-diagonal gluon condensation $\langle -\vec{X}_\rho \vec{X}_\rho \rangle$ 
also generates a mass term for $\vec{B}_\mu$.
In $SU(2)$ this produces a kinetic term for $\hat{n}$ 
in the Faddeev-Skyrme model. For arbitrary $SU(N)$ the equivalent expression is
\begin{eqnarray}
\lefteqn{\Lambda^2 g^2\vec{B}_\mu \cdot \vec{B}_\mu} \nonumber \\ 
&=& \Lambda^2 g\vec{B}_\mu \cdot \partial_\mu (\hat{n}_i \times \hat{n}_i) \nonumber \\
&=& \Lambda^2 (\partial_\mu \hat{n}_i)^2,
\end{eqnarray}
where $\Lambda$ is determined by $\langle -\vec{X}_\rho \vec{X}_\rho \rangle$.
Hence the general form of the Faddeev-Skyrme model extends easily to higher $N$.

\section{Discussion} \label{sec:discuss}
It is useful to express the QCD Lagrangian using the roots of representation theory.
Highlighting the group structure in this way greatly facilitates the extension of 
low $N$ results to higher $N$. So long as Abelian dominance holds, as expected at
low-energy, high $N$ results can be read off after knowing the two-colour result so long
as the quartic cross-terms remain insignificant. This is true at low loop order and 
seems reasonable at the qualitative level otherwise.

The extension of the energy eigenvalue analysis to higher $N$ has been doubted by
some people, but here follows easily from the $N=2$ case. Of course the imaginary part
also follows but various stability arguments also generalise rather well.
The Faddeev-Skyrme model has also been seen to generalise in an intuitive way to higher $N$.

\begin{ack}
The author greatly appreciates K.-I.~Kondo for many useful discussions and readings of the
manuscript, as well as for his assistance in subsection \ref{subsec:eigenvalue}.
\end{ack}


\end{document}